# Software Defined Edge Computing for Distributed Management and Scalable Control in IoT Multinetworks

Di Wu[1], Xiang Nie[1], Hanhui Deng[1], and Zhijin Qin[2]
[1]Key Laboratory for Embedded and Network Computing of Hunan Province, Hunan University, China
[2]School of Electronic Engineering and Computer Science, Queen Mary University of London, UK

**Abstract**: Recently, edge computing has emerged as a promising paradigm to support mobile access in IoT multinetworks. However, coexistence of heterogeneous wireless communication schemes brings about new challenges to the mobility management and access control in such networks. To maintain its computing continuum, Software-Defined Networking (SDN) architecture can be applied to the distributed edge networks, which provides consistent frameworks for mobility management and creates secure mechanisms for access control. In this paper, we first review the state-of-the-art SDN technology and its applications in edge computing. In particular, we highlight the prominent issues related to mobility management and access control for SDN in the edge computing environment. We also introduce a number of effective solutions to these issues by presenting real-world testbed experiments, prototypes and numerical analysis of adopting SDN based edge computing paradigms for IoT multinetworks.

## 1. Introduction

Urban-scale Internet of Things (IoT) applications in smart cities have the potential to generate a huge amount of data. As reported by Cisco, the connected IoT mobile devices (MDs) will exceed 11.6 billion and the global mobile traffic will be more than 49 exabytes per month by 2021 [1]. Such large amount of data brings huge pressure on the core networking backbone. Recently, edge computing paradigms are proposed to ease the workload of the core system [2] by distributing partial computing services and resources to the network edge. However, MDs connected to the edge network are usually configured with heterogeneous radio interfaces and access the networks in a location-dependent manner. Such geographical IoT multinetwork [3] structure brings potential issues in maintaining the computing continuum of the MDs and the network, such as mobility management and access control.

As a successful solution to ease the network management, Software Defined Network (SDN) is a promising solution to provide the computing continuum to the edge based IoT multinetworks. However, the solutions to distribute SDN technologies to the edge are not well explored to accommodate the requirements of the current network. In this paper, we aim to provide an overview of the existing solutions of applying SDN to edge computing, identify the challenges of developing well-suited SDN technologies in the edge network, and propose some feasible solutions.

## 2. Recent Development of SDN Architecture

**Software Defined Network:** In order to ease network management, the software defined networking (SDN) paradigm has been adopted to separate the control plane from the data plane. Particularly, the control plane makes decisions about traffic management, and the data plane involves mechanisms of forwarding traffic. The controller is adopted to manage the behaviors of the networks. By doing so, the flow based routing becomes quick and flexible, and rescheduling over the network components becomes possible as well. Most SDN control modes are centralized, in which controllers are deployed in the cloud server and all routes are decided by the controller based on a global network view. These SDN controllers focus on improving performance of single instance, such as network capacity optimization, which generally supports OpenFlow [4] as the standard southbound protocol. Generally, such centralized scheme suffers from the single point of failure problem and has serious implications for scalability and reliability due to the lack of fault tolerance mechanisms. Nevertheless, SDN still presents many new features, such as the logical centralized control model, unprecedented programmability, and flow-based model, which have the potential to fit into the highly scalable networks.

**Scalable SDN:** In IoT networks, the vast amounts of data flows generated by the massive number of MDs may bring huge pressure to the centralized SDN architecture. Hence, building scalable SDN systems is essential to provide scalability while preserving the desirable features of SDN. Some popular approaches have been proposed to scale SDN recently. Specifically, offloading partial workload from controllers to switches has been proposed to reduce the load of the centralized controller. However, this approach requires modifications of the switch hardware, thus limiting its usage to some scenarios of specific applications. An alternative approach is to design a distributed SDN system with multiple control planes, where the whole network view is distributed among multiple controller instances. This distributed method is very suitable for edge computing paradigms in IoT networks, as it can achieve excellent scalability by managing multiple network partitions independently. Moreover, each control plane can exchange network information and perform interoperation through the standard east/westbound interfaces. The availability and scalability issues in distributed SDN has been discussed in [5]. There are also very popular and well-known open source projects including OpenDaylight (ODL) and Open Network Operating System (ONOS), in which the adopted controllers implement an east/westbound protocol, e.g. SDNi, to enable communications between controllers.

**Push SDN to Edge:** The wide adoption of edge computing as well as the potential of deploying SDN in distributed edge networks are gradually becoming the next important topic to explore. It transforms the centralized processing capabilities of traditional SDN into a distributed architecture. Such architecture can push various processing tasks from the center of network to the edge. Moreover, it sets up additional nodes that sit down between the cloud servers and the MDs for edge processing purposes. By using such an architecture, various tasks can be offloaded from the cloud servers to the edge nodes, and data generated by the MDs can be cached on the edge nodes as well. As a result, the latency of critical applications is reduced, the reliance on the cloud can be mitigated, and the vast amounts of data generated by IoT devices can be managed in an agile SDN-Edge-IoT ecosystem. To maintain the capabilities of SDN on the edge nodes, a distributed control plane is more than desired in order to manage the mobile access of IoT devices. The common method is to build distributed SDN controllers into a controller cluster. For example, a distributed SDN architecture for IoT through the integration of mobile edge computing systems has been presented in [6], in which controllers are deployed on the edge servers to manage the MDs within its partition. By adopting such a dispersion, the integrated SDN and edge computing presents a feasible solution for mobile access in IoT multinetworks.

# 3. State-of-the-Art Solutions and Prominent Issues

## State-of-the-Art Distributed SDN

In distributed SDN, architecture, mobile access, flow scheduling, and security are essential to integrate SDN with edge computing and IoT multinetworks. The related work is summarized as follows.

**Architecture Design:** Currently, centralized controllers are mainly adopted in enterprise data center networks (DCN), such as Google B4 network. A distributed method has been proposed in [7] to improve the scalability of DCN by modifying the hardware in data plane. Another distributed SDN architecture is to distribute the network view among multiple controller instances [5].

**Mobile Access:** To address SDN access in heterogeneous wireless networks, Vertical handover based on OpenFlow has been investigated in the GENI testbed [8]. Mechanisms such as SoftCell [9] also use SDN to specifically handle mobility and handover to support scalable and flexible access of cellular core network.

**Flow Scheduling:** A scheme has been prototyped in MINA [3] to realize flow scheduling for video data between WiFi and Bluetooth networks. Additionally, in combination with mobility management, flow scheduling of IoT on the edge has been investigated in [10]. Moreover, contextual and flow-based access control has also been studied in [11] by using scalable host-based SDN techniques.

**Security:** For secured access in mobile scenario, FreeSurf introduces an authentication method using SDN for local wireless access control [12]. More recently, device fingerprint has been proposed to secure IoT systems in [13], where the corresponding location authentication mechanism has also been designed by adopting physical layer and other signatures.

## Issues in Distributed SDN for Edge Computing

At the time of writing, the design of SDN systems is still in its infancy in terms of architectures, schemes, and mechanisms. The prominent issues are discussed as follow, with particular focus on mobility management and access control.

**Hybrid Architecture to integrate SDN and Edge Computing:** The distributed SDN should scale to accommodate the needs for the large number of MDs in the IoT multinetworks. Besides, current southbound or east/westbound protocols are insufficient to support communications between the network components, such as switch and MDs in the heterogeneous networks. Therefore, *it is essential to design a hybrid architecture to integrate SDN functionalities with edge computing by considering the scalability, heterogeneity and compatibility of the networks.*

**Mobility Management in Distributed Networks:** When IoT devices roam frequently at the network edge, each controller can build up a mobility map of these devices. In order to manage the software defined edge computing, it is required to combine scalable control with consistent mechanisms. As a result, the distributed SDN can coordinate multiple controllers and switches for information exchange in supporting seamless handover, scalable control and fault tolerance in the networks. Therefore, *deploying mechanisms into distributed SDN for coordinating mobility of devices across distributed networks is essential to ensure the operation of the SDN on the edge.*

**Flow Scheduling based on Network Partition:** In large-scale networks, SDN controllers could divide the network into multiple partitions according to the geographic areas, flow requests, and mobile information. Each controller should be able to handle data service, resource allocation and flow scheduling within it individual partition so as to satisfy the individual flow request from IoT devices with guarantee on optimal performance of the entire system. Therefore, *the partition methods adopted by distributed SDN for edge computing paradigms should take flow scheduling into consideration and deliver optimal performance based on the flow requirements.*

**Secured Mobile Access:** Due to the lack of proper authentication and authorization schemes, the network attackers can easily find the design faults to discover security key and intercept the important information in software defined edge computing networks. This leads that the relevant components deployed on the network edge, such as IoT devices, access points (APs), controllers, become vulnerable to malicious attacks (e.g. replay attack, Sybil attack, wormhole attack). Therefore, *it is more than desired to build a feasible and secured coordination scheme for SDN controllers and networks edge devices.*

## 4. Enabling Software Defined Edge Computing for IoT

In the following, we propose a SDN-Edge-IoT ecosystem architecture for IoT multinetworks to properly address various design issues, including mobility management, network availability, scalability, and security.

### SDN-Edge-IoT Ecosystem Architecture

To efficiently tackle the challenges and issues in the integration of SDN and edge computing, we propose an SDN-Edge-IoT ecosystem architecture built on the OpenFlow framework, as illustrated in Fig. 1. In this architecture, the controllers are distributed in proximity to the network edge. The entire network is divided into several partitions to present different geographical areas. At the network edge, each IoT device is associated with heterogeneous APs based on the needs of flow types. The OpenFlow switches are connected with SDN controllers to provide high speed services with low latency. Because of the heterogeneity of APs and MDs, the southbound interface protocols are extended with middleware applications to handle network compatibility, so as to enable agile access and handover.

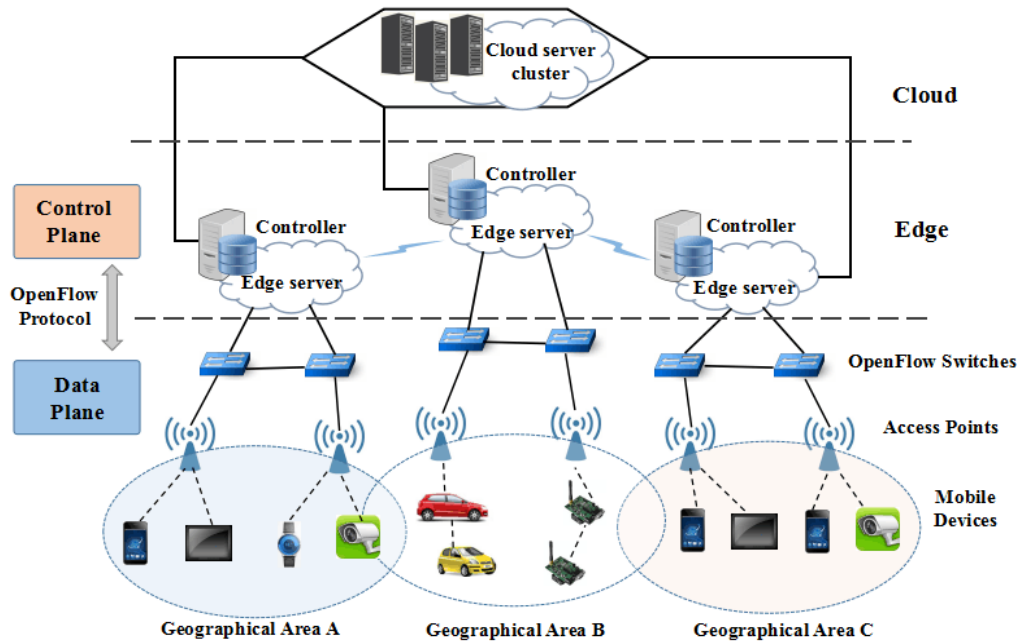

**Figure 1.** An SDN-Edge-IoT ecosystem architecture.

In the control plane, as each controller is responsible for a specific geographical area, it will have a local view of the network status. Information pertaining to service requests and flow transmissions are managed by the partition-dependent controllers. The flow requests and resource allocation are determined through coordinating these distributed controllers. Through the extension of east/westbound interface to support communications between controllers, distributed mobility management and access control can be achieved by exchange the information recorded by different controllers via connected switches. Therefore, above SDN-Edge-IoT ecosystem architecture is able to meet the network availability and scalability needs in software defined Edge computing.

## Distributed Mobility Management

When IoT devices roam between different partitions, a consistent controller coordination scheme is required to manage the mobility of the devices. The distributed SDN systems can ensure consistency of controllers based on the exchanged information through east/westbound interface. However, a fundamental problem of maintaining consistency is to locate the controllers that are responsible for specific MDs in the network. If each controller maintains a database containing the information of all other controllers, the controller information update would introduce significant overhead when controllers join or leave the SDN system. Therefore, simple and efficient methods are required to maintain consistency in mobility management.

In order to realize the desired mobility management, we propose a distributed lookup protocol to address the mobility management problem, based on the distributed hash table (DHT) of Chord [14] structure, and its lookup scheme using finger table. As shown in Fig. 2a, the proposed method maintains a controller network by using an ordered ring overlay, where the distributed controllers act as overlay nodes in the consistent hashing framework.

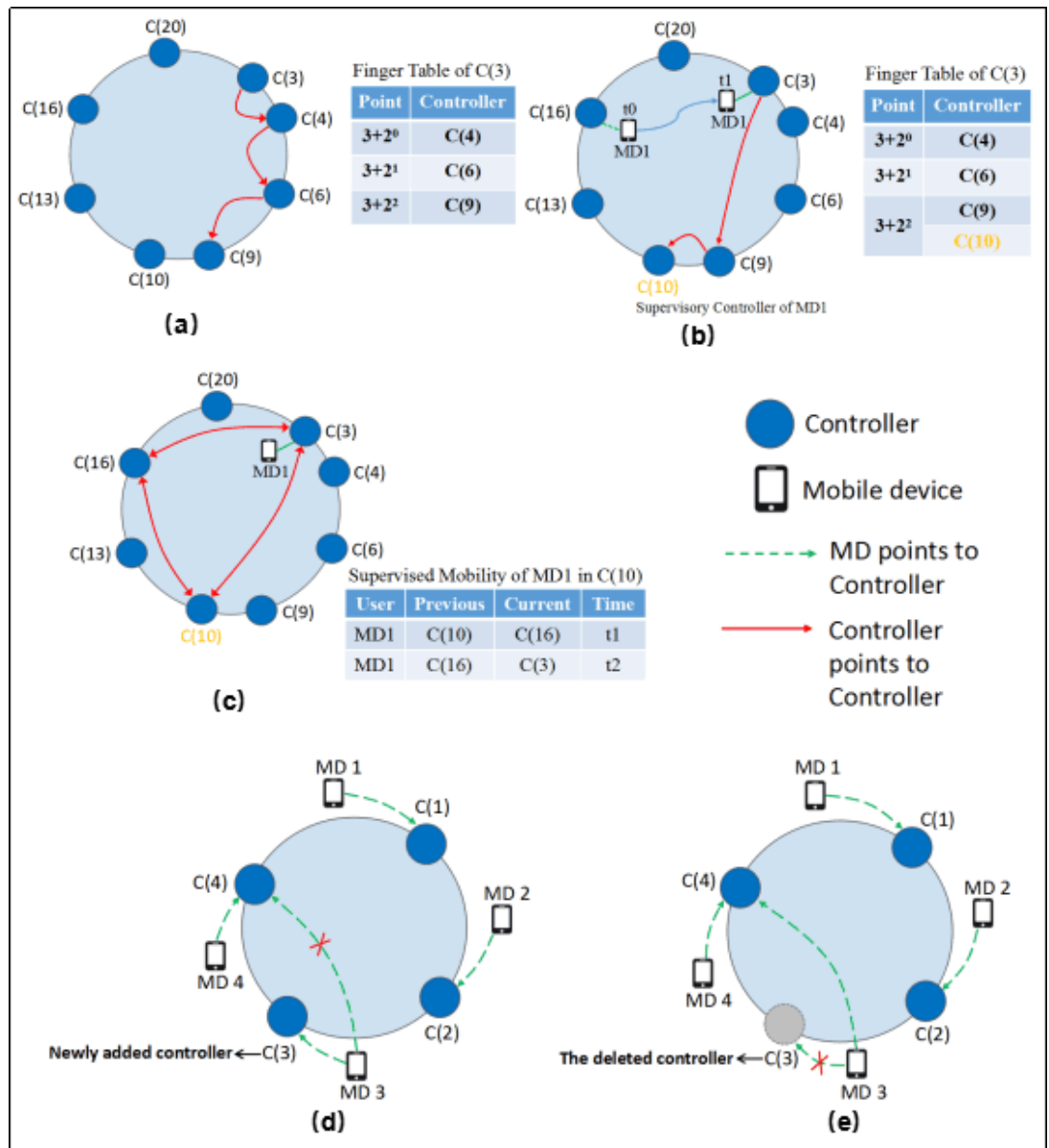

**Figure 2.** Mobility management in the considered overlay networks: (a) overlay case, (b) mobility case, (c) handover case, (d) add a controller, and (e) delete a controller.

As a bootstrapping step, when there is a new IoT MD joins the distributed SDN network, a supervisory controller will be its initially associated one. In the mobility case illustrated in Fig. 2, the mobile ID of each IoT MD is hashed by the overlay structure. Then an integer key is obtained to localize the IoT MD's supervisory controller. Since every controller can use consistent hashing to localize the supervisory controller of any IoT MD, the supervisory controller is used to track the previous and current associated controllers for the newly joined IoT MD. As illustrated in Fig. 2(b) and Fig. 2(c), once

MD M1 moves into controller C(3)'s partition, the new associated controller C(3) can localize its supervisory controller C(10) and learn that C(16) is the previous associated controller for M1. After this, C(3) can communicate with C(16) directly to fetch the previous session between M1 and C(16) and reroute flows to its current partition.

In addition, we propose a Personal AP protocol running in the controller for mobility management between IoT MDs and APs. Particularly, the proposed Personal AP protocol replicates the complete association parameters[1] between each pair of MD and AP, and reinstates the association parameters at the AP, which is best-connected to the MD. Then MDs perceive the newly connected AP as its old AP. By doing so, the MAC layer management information keeps intact without requiring the re-establishing of association states between the MD and the network backend.

## Scalable Control and Fault Tolerance

The key of implementing scalable mobility management is to maintaining the consistency of controller clusters. The joining and leaving of controllers are two typical operations to achieve scalable control in distributed scenarios. As illustrated in Fig. 2(d) and Fig. 2(e), when there is a controller that newly join in the existing overlay, its successor is identified first by performing a lookup in the overlay structure based on the controller ID. Once the successor is located, the newly joined controller selects the key of the successor. Then the newly joint controller updates its successor's former predecessor as its predecessor, and update its predecessor according to that of its successor. On the contrary, when a controller decides to leave an existing overlay, it first move all keys controlled by the controller to its successor. Afterwards, the controller updates its successor's predecessor to that of itself, and sets its predecessor's successor to its successor. It is worth noting that before the controller leaves, the default rule is to copy the relevant control information , such as network status and flow status, maintained by the controller to its successor for consistency purposes.

Additionally, fault tolerance is another critical issue in scalable control. We need to tackle failures of different components in the distributed software defined edge computing architecture. As for controller failure, robust control can be achieved by replicating the data of the faulty controller to its successors. Specifically, if there are failures in the local controller, a new successor is adopted to replace the failed one. While for the finger-table failures, alternate paths are selected adaptively while routing. If there is no response from a finger, the previous fingers in the local table or finger-table replicas from one successor will be adopted. While for the AP failure, the associated controller can detect it first, and then flows going through the failed AP will be redirected to the other APs in its partition. Thereafter, IoT devices can associate with the newly assigned AP.

---

[1] Here, the association parameters contain state information, including the MD MAC address, AP MAC address, MD association ID, data frame sequence number, security keys, and flow status.

## Flow Scheduling

If there are flow requirements from IoT devices, the SDN controller will find available resources on the network edge to satisfy the flow requests and achieve the optimal performance of the whole partition. However, the IoT flow requirements are usually depicted in an abstract manner. As a result, they are irrelevant to the underlying network and device resources specifications. Therefore, in order to schedule data flow efficiently, the SDN controller of each partition should build up a partition view. As illustrated in Fig. 3, the partition-dependent IoT MDs associate with heterogeneous APs and request various types of data flow from the corresponding data server. As these devices roam across different partitions, the records of the new and left IoT MDs are maintained in the local controller to indicate resource usage and user density. The Network Calculus method [3] has been proposed for the controllers to get the partition view by obtaining the up-to-date status of flow scheduling within its partition.

After the partition view of current network status is available, the controller can manage handover between heterogeneous access networks. Particularly, based on the current capacity of the controlled partition, the supported radio access technologies and the types of requested services, the controller will assign the newly joined IoT MD to the optimal AP. To select the best available APs for dynamic IoT flow requirements under multiple network constraints, we could formulate it as a generalized assignment problem (GAP) and use greedy heuristic methods [10] to match flow requests with optimal or near-optimal resource allocation.

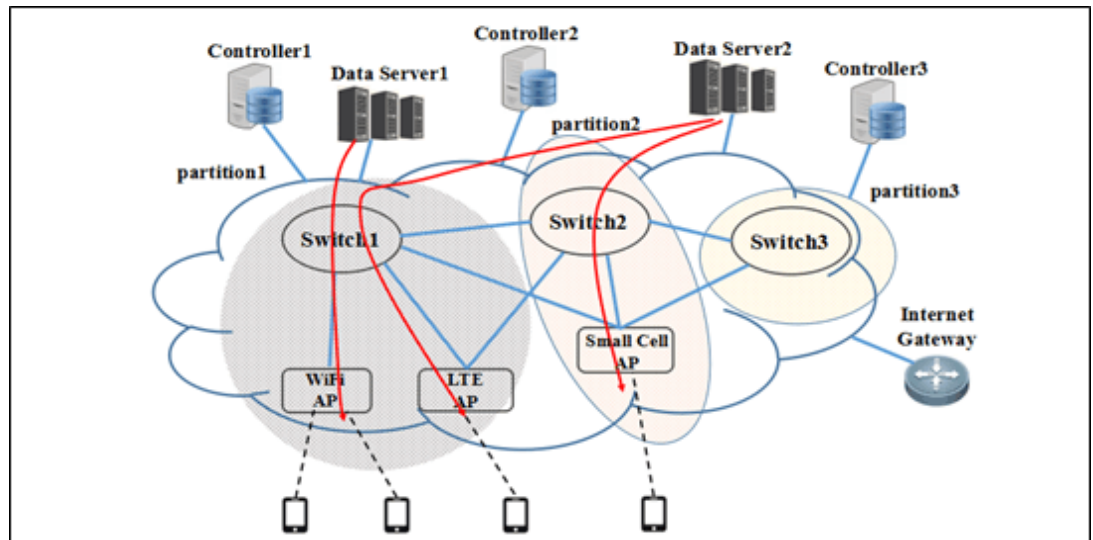

**Figure 3.** A heterogeneous network space based partition.

## Secured Mobile Access

The security is critical for the development of distributed SDN for edge computing, especially in IoT multinetworks where security mechanisms are limited for IoT devices in mobile scenario. The examples in [15] show that edge computing system obviously faces many security and privacy threats. Compared to centralized systems, the distributed features of edge computing are more vulnerable to threats from malicious attacks. Therefore, the SDN control plane needs to be actively involved to mitigate the existing security problems. The controller should defend against not only the traditional network attacks, such as Dos/DDoS, but also the injection of malicious/false flow rules against edge computing. Specifically, since controllers deployed in our system split the edge network into several partitions according to the represented geographical areas, the location-based authentication is of great interest to the network access control of IoT multinetworks.

The location group is defined as a set of APs from the IoT multinetworks in the partition, whose wireless coverages share the access-granted area. The location group information of each partition is regarded as one type of network information, which is recorded and maintained in the corresponding controller. Fig. 4(a) illustrates two access-granted areas as marked by the shaded areas of two AP location groups, G1={AP1, AP2} and G2={AP3, AP4}. For location authentication procedure as shown in Fig. 4(b), the controller first distributes information including location group membership and security key to the APs. Meanwhile, the key information is updated periodically. The APs will then broadcast the received keys to the IoT devices in the partition by beacon messages. The location of an IoT MD can be determined by those IoT devices that collect and present all the key information from the corresponding APs.

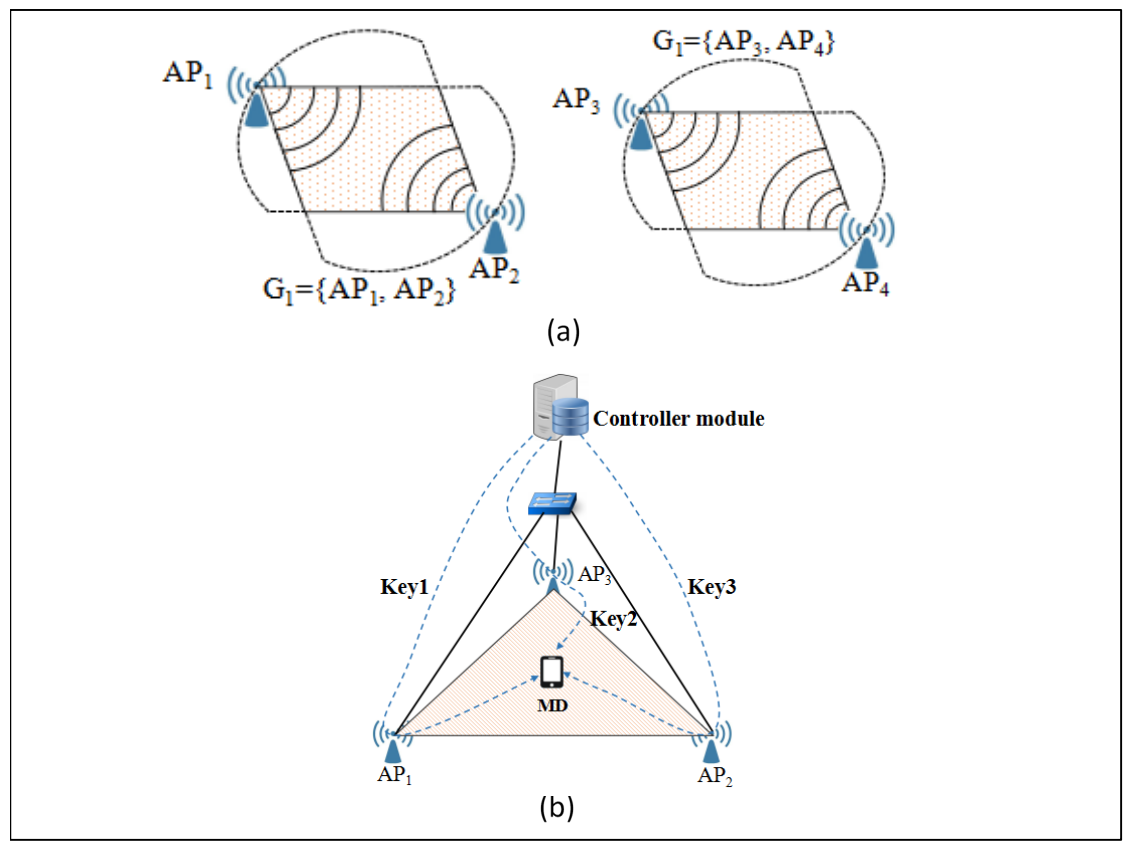

**Figure 4.** Authentication within access-granted network area: (a) access-granted areas, and (b) authentication procedure.

# 5. Results and Analysis from Experiment Testbed

In this section, we implement a prototype system called LEDGE to test the feasibility of SDN-Edge-IoT systems. The well-known ORBIT is used as our wireless network testbed, which contains components such as WiFi, WiMAX, and USRP2, and can be accessed by the management framework.

## Analysis of Mobile Access

As real mobile access patterns of IoT devices on the edge of large multinetworks follow event or motivation driven behaviors, we chose a school building to collect mobile traces based on the classes. During the experiment period, students have different behaviors, such as coming to the building, leaving the building, and staying in the building. Accordingly, the statuses of MDs are classified as "joining", "leaving", and "staying". Particularly, there are 8 OpenFlow-enabled ORBIT nodes deployed in the building as corresponding APs. Two SDN controllers are used to construct a distributed scenario to schedule various requests from more than 300 IoT devices. We also compare the performance in terms of throughput (flow) and delay of LEDGE with two OpenFlow-based wireless access control schemes proposed by FreeSurf [12] and MINA [3].

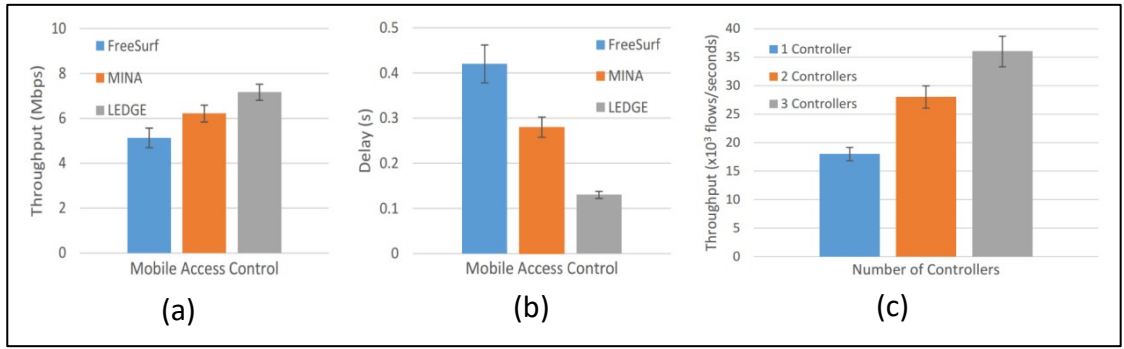

**Figure 5.** Performance of mobile access: (a) mobile throughput, (b) mobile delay, and (c) impact of scalability on throughput.

From Fig. 5(a) and Fig. 5(b), the observed performance indicates that LEDGE outperforms FreeSurf and MINA both in terms of throughput and delay when MDs are located in the access granted areas. LEDGE can achieve the average throughput of 7.16 Mbps, while MINA can offer 6.21 Mbps throughput and FreeSurf only reaches 5.13 Mbps. The average delay in LEDGE is about 0.13 s, while the delay in MINA is about 0.28 s and FreeSurf is 0.42 s. The performance of LEDGE is better than both MINA and FreeSurf because LEDGE adopts agile mobility management and access control on the network edge.

We also verify the scalability of LEDGE in the ORBIT testbed using three ORBIT sandboxes. If there is a flow, a Packet-In message will be sent to the controller. Therefore, the Packet-In message is used to evaluate the scalability of flow scheduling in LEDGE. Fig. 5(c) demonstrates the throughput of LEDGE with varying number of controllers. It is observed that increasing the number of controller results in an increment of throughput in an almost linear manner. This is due the fact that each controller in LEDGE can mainly control the traffic flows within its own partition.

## Analysis of Location Authentication

An ORBIT sandbox with three AP nodes is chosen in the second experiment. Fig. 6(a) illustrates mobile scenarios, in which AP nodes 1, 2, 3 form a location group $AP_1$, $AP_2$, $AP_3$, and the access-granted area is defined by the triangular area marked with dashed blue lines. These three APs connect with an OpenFlow switch (labeled as node 4), which is also connected with a SDN (node 5) and a MD (node 6). All wireless links have 11Mbps data rate. The controller periodically generates and distributes individual keys Key1, Key2, Key3 to the corresponding APs. Then the APs broadcast the keys in their beacon messages to MDs.

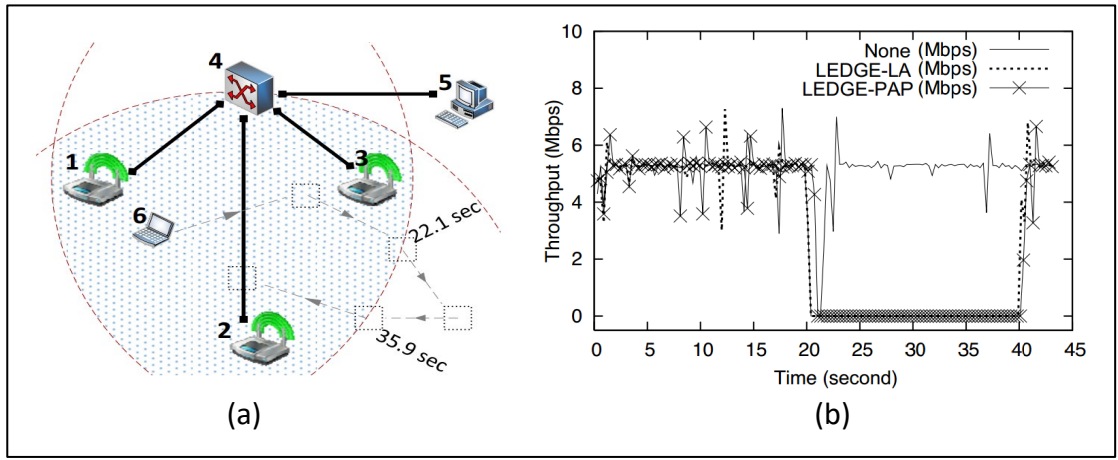

**Figure 6.** Performance of location authentication: (a) mobile scenario, and (b) mobile throughput.

In our location authentication experiments, the beacon period is set as 100 ms in order to guarantee prompt delivery of the keys to the MD in the access-granted areas. In the experiment, one TCP stream is directed from the MD (node 6) to the SDN controller (node 5). We create a mobility pattern that node 6 moves out of the access granted area at 22.1s and return to the area at 35.9s. In the comparison, the data traffic tests are carried out in three settings: 1) no access control mechanism (None); 2) with location authentication (LEDGE-LA) mechanism; and 3) with Personal AP protocol (LEDGE-PAP).

Fig. 6(b) illustrates the throughput of the TCP connection with and without LEDGE operations respectively. All TCP streams are disrupted at 22.1s for both scenarios with or without LEDGE operations since the MD (node 6) loses the connection with associated AP (node 1). The TCP stream without network access control mechanism is able to come back to the normal throughput rapidly after associating with another new AP (node 3). However, the TCP streams with LEDGE-LA and LEDGE-PAP cannot recover their throughput until the MD returned to the access-granted area at time 35.9s. It is worth noting that the TCP streams with LEDGELA and LEDGE-PAP have a 4 second lag before recovering to their normal throughput, which is caused by the congestion and flow control timing mechanisms adopted by TCP. Both LEDGE-LA and LEDGE-PAP achieve the same access control goal within one location group though with protocol operations.

# 6. Conclusion

SDN and edge computing have recently emerged as a promising paradigm to support mobile access in IoT multinetworks. These communication and networking technologies also bring new challenges in both design and deployment of software defined edge computing networks. To meet the network availability and scalability needs, we have presented an SDN-Edge-IoT ecosystem architecture for mobility management and access control in IoT multinetworks. Relevant scalability and consistency issues to maintain the distributed system have also been carefully addressed. In addition, security has been integrated into the design of protocols in the communication architectures. The results collected from real testbed confirm that our software defined edge computing system is able to provide agile mobility management and access control for IoT multinetworks.